\documentclass[12pt]{article}
\usepackage[T1]{fontenc}
\usepackage{cite}
\usepackage{mathrsfs}
\usepackage{amsmath}
\usepackage{amssymb}
\usepackage[dvips]{graphicx}
\usepackage{tabularx}
\usepackage{bm}

\newenvironment{minilinespace}{\baselineskip = 8mm}{}
\begin{document}

\begin{titlepage}

\begin{flushright}
{
	KUNS-2083, WU-AP/269/07
}
\end{flushright}
\vspace{1cm}

\begin{minilinespace}
\begin{center}
	{\Large
		{\bf 
			Hawking radiation of a vector field\\
			and gravitational anomalies
		}
	}
\end{center}
\end{minilinespace}
\vspace{1cm}

\begin{center}
Keiju Murata$^{a,1}$ and Umpei Miyamoto$^{b,2}$\\
\vspace{.5cm}
{\small {\textit{$^{a}$
Department of Physics, Kyoto University, Kyoto 606-8501, Japan
}}
}
\\
\vspace{5mm}
{\small \textit{$^{b}$
Department of Physics,
 Waseda University, Okubo 3-4-1, Tokyo 169-8555, Japan
}}
\\
\vspace*{1.0cm}

{\small
{\tt{
\noindent
$^{1}$ murata@tap.scphys.kyoto-u.ac.jp
\\
$^{2}$ umpei@gravity.phys.waseda.ac.jp
}}
}
\end{center}

\vspace*{1.0cm}



\begin{abstract}

Recently, the relation between Hawking radiation and gravitational
 anomalies has been
 used to estimate the flux of Hawking radiation for a large class of black objects.
In this paper, we extend the formalism, originally proposed by Robinson and Wilczek, to the Hawking radiation of vector particles (photons).
It is explicitly shown, with Hamiltonian formalism, that the theory of an electromagnetic field on $d$-dimensional spherical black holes reduces to one of an infinite number of massive complex scalar fields on 2-dimensional spacetime, for which the usual anomaly-cancellation method is available. It is found that the total energy emitted from the horizon for the electromagnetic field is just ($d-2$) times as that for a scalar field.
The results support the picture that Hawking radiation can be regarded as an anomaly eliminator on horizons. Possible extensions and applications of the analysis are discussed.
\end{abstract}

\end{titlepage}

\section{Introduction}

Understanding the physics of black hole horizons, 
such as black hole entropy and Hawking radiation, 
should hint at how we can construct
a quantum theory of gravity. 
Recently, there has been progress
in understanding the black hole entropy~\cite{Carlip:2006fm}. 
The breakdown of a diffeomorphism symmetry at a horizon,
namely an anomaly, has been found to play an important role.   
Since Hawking radiation~\cite{Hawking:1974sw} as well as entropy 
is a property inherent in horizons, it is natural to expect that Hawking radiation is also associated with anomalies. 

Many years ago, Christensen and Fulling found that Hawking radiation can be derived from the trace anomaly \cite{Christensen:1977jc} for ($1+1$)-dimensional Schwarzschild black holes. 
In their approach, as usual, boundary conditions both
at the horizon and infinity are required to specify the vacuum.
Hence, it seems difficult to attribute Hawking radiation to a property of the event horizon.
It should be also mentioned that the method is not applicable to more
than 2-dimensions. 
Recently,  Robinson and Wilczek have suggested a new derivation of Hawking
radiation from Schwarzschild black holes through gravitational
anomalies~\cite{Robinson:2005pd}.
The remarkable point is that the derivation is applicable to any number
of dimensions.
In their argument, Hawking radiation is a compensating flux canceling the gravitational anomalies at the horizon.
The advantage of the derivation is that it requires information only at the horizon.
Furthermore, Iso et al.~showed that the Hawking radiation 
from a Reissner-Nordstr\"{o}m black hole can be explained as the fluxes canceling
the gravitational and $U(1)$-gauge anomalies~\cite{Iso:2006wa}.
They also clarified the boundary condition at the horizon. 
Their technique was also applied to rotating black
holes~\cite{Iso:2006ut,Murata:2006pt,Setare:2006hq,Xu:2006tq,Iso:2006xj,Jiang:2007gc}. 
The angular-momentum flux from rotating black holes, which is regarded as a $U(1)$ current flow,
can be calculated as well as the energy flux.
Applications for various black holes are found in
\cite{Vagenas:2006qb,Jiang:2007pn,Jiang:2007wj,Kui:2007dy,Shin:2007gz,Peng:2007pk,Jiang:2007mi,Das:2007ru,Jiang:2007pe}. 
Very recently, the extension to black rings, which have
a horizon topology of $S^1\times S^2$, was done in~\cite{Chen:2007pp,Miyamoto:2007ue}. 
The thermal distribution of Hawking radiation has also been derived from
the anomaly viewpoint~\cite{Iso:2007kt,Iso:2007hd}. 

Although much work has been devoted to this theme,
these analyses are focused on scalar particle radiations. 
Therefore, to verify the universality of the anomaly cancellation
technique, Hawking radiation for other particles with non-zero spins
should be investigated. 
In this paper, we extend the anomaly cancellation method to the vector
particle radiation from generic spherically symmetric black holes.

The organization of this paper is as follows.
In Sec.~\ref{sec:EMonSSS}, we investigate the general properties of an
electromagnetic field on spherically symmetric spacetimes. In
particular, we show that the electromagnetic field on the spherical
spacetimes is equivalent to a set of infinite numbers of complex scalar
fields on the ($t, r$) sector of spacetimes, where $r$ is the radial
coordinate. In Sec.~\ref{sec:HRandGA}, using the results obtained
in the previous section, we apply the anomaly cancellation method to the
photon emission from spherically symmetric black holes.
The final section is devoted to a summary and discussion.
We use units in which $c=G=\hbar=k_B=1$ throughout this paper.

\section{Electromagnetic fields on spherically symmetric spacetimes}
\label{sec:EMonSSS}

\subsection{ Dimensional reduction of electromagnetic field theory }

Let us consider an electromagnetic field on $d$-dimensional spherically
symmetric static spacetimes, of which the line element is given by 
\begin{equation}
\begin{split}
 ds^2 =& g^{(d)}_{\mu\nu}dx^\mu dx^\nu \\
=& -f(r)dt^2 + f(r)^{-1}dr^2 + r^2 d\Omega_{n}^2\ ,
\end{split}
\end{equation}
where $n\equiv d-2$ and $d\Omega_n^2$ is the line element of $S^{n}$. 
$f(r)$ is a function admitting an event horizon, $r=r_H$, where $f(r_H)=0$.
The surface gravity is given by $\kappa=f'(r_H)/2$.
For the 2-dimensional covariance to be apparent, we also use the expression of the line element given by 
\begin{equation}
 ds^2 = g_{ab}(x^a)dx^a dx^b + e^{-2\phi(x^a)}d\Omega_{n}^2\ ,
\end{equation}
where $a,b = t,r$. 
The action for the electromagnetic field is
\begin{equation}
\begin{split}
 S &= -\frac{1}{4}\int d^d x \sqrt{-g^{(d)}} F_{\mu\nu}F^{\mu\nu}\\
  &= -\frac{1}{4}\int d^2 x \sqrt{-g}\,e^{-n\phi}\int d\Omega_n
     [F_{ab}F^{ab} + 2F_{ai}F^{ai} + F_{ij}F^{ij}]\ ,
\end{split}
\label{eq:action1}
\end{equation}
where $x^i$ is the coordinates on $S^{n}$.
Note that this action is invariant under the gauge transformation,
\begin{equation}
 A_\mu(x^\mu) \rightarrow A_\mu(x^\mu) - \partial_\mu \lambda(x^\mu)\ , 
\end{equation}
where $\lambda$ is an arbitrary scalar function.

Now, we decompose the vector potential, $ A_\mu $, into vector and scalar modes
on $S^{n}$~\cite{Rubin:1984tc,Higuchi:1986wu,Kodama:2003jz,Ishibashi:2003ap}.
These two modes are decoupled from each other in the action, and one can treat them separately. 
The vector modes are decomposed as
\begin{equation}
 A_a(x^\mu) =0,\quad
 A_i(x^\mu) = \sum_N \sqrt{2}\,\alpha_N(x^a)\mathbb{V}^N_i(x^i)\ ,
\end{equation}
where $ \mathbb{V}^{N}_i $ is the $n$-dimensional vector spherical harmonic solving 
\begin{equation}
 (D_j D^j + k_V^2)\mathbb{V}^N_i = 0\ ,\quad D^i \mathbb{V}^N_i=0\ .
\end{equation}
Here, we denote the covariant derivative on $S^n$ by $D_i$.
The eigenvalue $k_V$ is given by $ k_V^2 = l(l+n-1)-1,\ (l=1,2,\cdots) $.
It should be noted that $l=0$ mode does not exist for the vector
harmonics. 
$N$ is the index running over all vector harmonics, which are normalized as
\begin{equation}
 \int d\Omega_n \mathbb{V}^{N}_i\mathbb{V}_{N'}^i{}^\ast = \delta_{NN'}\ .
\end{equation}
It is mentioned that since the gauge parameter $\lambda(x^\mu)$ contains no vector mode, expansion coefficient $\alpha_N(x^a)$ is gauge invariant.
Then, we can carry out the integration over the angular variables in the
action~(\ref{eq:action1}) to give
\begin{equation}
 S=-\sum_N \int d^2 x \sqrt{-g}\,e^{-(n-2)\phi}[\partial_a \alpha_N
  \partial^a \alpha_N^\ast + (k_V^2 + n-1)e^{2\phi}\alpha_N\alpha_N^\ast]\ .
  \label{eq:vector-action}
\end{equation}
From Eq.~(\ref{eq:vector-action}), we see that the vector modes are equivalent to a set of complex massive scalar fields, coupling to a dilaton, in two dimensions.

The scalar modes on $S^n$ are decomposed as
\begin{equation}
 A_a(x^\mu) = \sum_M A^M_a(x^a)\mathbb{S}^M(x^i)\ ,
 \quad
 A_i(x^\mu) = \sum_M \beta_M(x^a)D_i \mathbb{S}^M(x^i)\ ,
\end{equation}
where $\mathbb{S}^M$ is the $n$-dimensional scalar spherical
harmonics solving 
\begin{equation}
 (D_i D^i + k_S^2)\mathbb{S}^M = 0\ .
\end{equation} 
The eigenvalue $k_S$ is given by $k_S^2 = l(l+n-1),\ (l=0,1,2,\cdots)$. 
$M$ is the index running over all scalar harmonics, which are normalized as
\begin{equation}
 \int d\Omega_n \mathbb{S}_{M}\mathbb{S}_{M'}^\ast = \delta_{MM'}\ .
\end{equation}
Then, the action~(\ref{eq:action1}) becomes
\begin{equation}
 S = -\frac{1}{4}\sum_M \int d^2 x
  \sqrt{-g}\,e^{-n\phi}[F_M^{ab}F^M_{ab}{}^\ast
  + 2k_S^2 e^{2\phi}(A_M^a-\partial^a\beta_M) 
(A^M_a{}^\ast-\partial_a\beta_M^\ast)]\ ,
\label{eq:action_scalar1}
\end{equation}
where 
$F^M_{ab}\equiv\partial_a A^M_b-\partial_b A^M_a$. 
The gauge parameter $\lambda(x^\mu)$ is also decomposed into the scalar harmonics,
\begin{equation}
 \lambda(x^\mu) = \sum_M \lambda_M(x^a)\mathbb{S}^M(x^i)\ ,
\end{equation}
and the gauge transformations for $A_a^M$ and $\beta_M$ become
\begin{equation}
 A_a^M \rightarrow A_a^M - \partial_a
  \lambda_M\ ,\quad
 \beta_M \rightarrow \beta_M - \lambda_M\ .
\end{equation}
From these transformation properties, one can find a gauge-invariant variable,
\begin{equation}
 \mathcal{A}^M_a \equiv A^M_a-\partial_a \beta_M.
\end{equation}
Making use of this variable, we can write the action for the scalar modes, Eq.~(\ref{eq:action_scalar1}), as
\begin{equation}
 S = -\frac{1}{4}\sum_M \int d^2 x
  \sqrt{-g}\,e^{-n\phi}[\mathcal{F}_M^{ab}\mathcal{F}^M_{ab}{}^\ast
  + 2k_S^2 e^{2\phi}\mathcal{A}_M^a \mathcal{A}^M_a{}^\ast]\ ,
\label{eq:action_scalar2}
\end{equation}
where 
$\mathcal{F}^M_{ab}\equiv\partial_a \mathcal{A}^M_b-\partial_b \mathcal{A}^M_a$. 
The action for $l=0$ mode is
\begin{equation}
 S=-\frac{1}{4}\int d^2 x
  \sqrt{-g}\,e^{-n\phi}\mathcal{F}_0^{ab}\mathcal{F}^0_{ab}{}^\ast\ .
\end{equation}
This is the action for a 2-dimensional electromagnetic theory, which has no
degree of freedom. Therefore, we do not consider the $l=0$ mode hereafter.

Thus, one can regard Eq.~(\ref{eq:action_scalar2}) as the action
for 2-dimensional massive electromagnetic fields (i.e., Proca fields) coupling to a dilaton. 
Furthermore, in the next part of this section, we will see that this theory is equivalent to
one for an infinite number of complex scalar fields.

\subsection{Equivalence between electromagnetic and scalar field theories}

First, note that constraints subsist in action~(\ref{eq:action_scalar2}).
For the constraints to be apparent, it is convenient to adopt Hamiltonian formalism~\cite{Dirac}.
From action~(\ref{eq:action_scalar2}), one can read off a Lagrangian as
\begin{equation}
\begin{split}
 \mathcal{L} =& \sqrt{-g}\left[-\frac{1}{4}e^{-n\phi}\mathcal{F}^{ab}\mathcal{F}_{ab}^\ast
 -\frac{1}{2}k_S^2 e^{-(n-2)\phi}\mathcal{A}^a \mathcal{A}_a^\ast\right]\\
 =&
 \frac{1}{2}e^{-n\phi}(\dot{\mathcal{A}_r}-\mathcal{A}_t') 
(\dot{\mathcal{A}_r^\ast}-\mathcal{A}_t^{\prime \ast})
 + \frac{k_S^2
 e^{-(n-2)\phi}}{2f}\mathcal{A}_t\mathcal{A}_t^\ast -\frac{k_S^2
 e^{-(n-2)\phi}f}{2}\mathcal{A}_r\mathcal{A}_r^\ast\ ,
\end{split}
\end{equation}
where $^\cdot\equiv \partial/\partial t$ and $' \equiv \partial/\partial r$.
Here, we have used the explicit form of a 2-dimensional metric,
\begin{equation}
 g_{ab}dx^a dx^b = -f(r)dt^2+f(r)^{-1}dr^2\ ,
\label{eq:2Dmetric}
\end{equation}
in the second line and omitted the index $M$.
Since the $l=0$ mode is irrelevant as we mentioned before, $k_S^2>0$ can be assumed.

The conjugate momenta of $\mathcal{A}_a$ and $\mathcal{A}_a^\ast$ are given by
\begin{align}
  \pi^t =& \frac{\partial \mathcal{L}}{\partial \dot{\mathcal{A}_t^\ast}}
  =0\ ,\label{eq:consraint1}\\
  \pi^{t\,\ast} =& \frac{\partial \mathcal{L}}{\partial \dot{\mathcal{A}_t}}
  =0\ ,\label{eq:consraint2}\\
  \pi^r =& \frac{\partial \mathcal{L}}{\partial
  \dot{\mathcal{A}_r^\ast}}
  = \frac{e^{-n\phi}}{2}(\dot{\mathcal{A}_r}-\mathcal{A}_t')\ ,\\
  \pi^{r\,\ast} =& \frac{\partial \mathcal{L}}{\partial
  \dot{\mathcal{A}_r}}
  = \frac{e^{-n\phi}}{2}(\dot{\mathcal{A}_r^\ast}-\mathcal{A}_t^{\prime \ast})\ .
\end{align}
The above momenta satisfy the following canonical commutation relations,
\begin{equation}
\begin{split}
 &\{\mathcal{A}_t(t,r),\pi^t{}^\ast(t,r')\} = \{\mathcal{A}_t^\ast(t,r),\pi^t(t,r')\}\\
=\,&\{\mathcal{A}_r(t,r),\pi^r{}^\ast(t,r')\} =
  \{\mathcal{A}_r^\ast(t,r),\pi^r(t,r')\} = \delta(r-r')\ ,
\end{split}
\end{equation}
where $\{\ ,\ \}$ represents a Poisson bracket. 
Equations~(\ref{eq:consraint1}) and (\ref{eq:consraint2}) are primary
constraints. Now, we define a whole phase space 
$M\equiv \{A_a,\pi^a,A_a^\ast,\pi^{a\,\ast}\}$ and 
a subspace which satisfies the primary
constraints, 
$M_0\equiv \{x\in M|\pi^t=\pi^{t\,\ast}=0\}$. 
Because of the primary constraints $\pi^t=\pi^t{}^\ast=0$, 
$\dot{\mathcal{A}}_t$ and $\dot{\mathcal{A}}_t{}^\ast$ cannot be written
in terms of the canonical momenta. 
However, we can write down the Hamiltonian as
\begin{equation}
\begin{split}
 \mathcal{H}_0 =& \pi^t \dot{\mathcal{A}_t^\ast} + \pi^{t\,\ast} \dot{\mathcal{A}_t} + 
 \pi^r \dot{\mathcal{A}_r^\ast} + \pi^{r\,\ast} \dot{\mathcal{A}_r} - \mathcal{L}\\
 =& 2e^{n\phi}\pi^r \pi^r{}^\ast + \pi^r \mathcal{A}_t^{\prime \ast} + \pi^r{}^\ast \mathcal{A}_t'
 -\frac{k_S^2 e^{-(n-2)\phi}}{2f}\mathcal{A}_t \mathcal{A}_t^\ast + \frac{k_S^2
 e^{-(n-2)\phi}f}{2}\mathcal{A}_r \mathcal{A}_r^\ast\ .
\end{split}
\end{equation}
In the second line, we have used the constraint equations
$\pi^t=\pi^{t\,\ast}=0$, and 
the $\mathcal{H}_0$ can be used only in $M_0$. 
To construct a Hamiltonian, $\mathcal{H}$, in the whole phase space $M$, 
$\mathcal{H}|_{M_0}=\mathcal{H}_0$ should be required, which is realized by
\begin{equation}
 \mathcal{H} = \mathcal{H}_0 + \mu^\ast \pi^t +  \mu
  \pi^{t\,\ast}\ .
\label{eq:H_tot}
\end{equation} 
Here, $\mu$ and $\mu^\ast$ are functions of the canonical variables.

Because the primary constraint $\pi^t = 0$
must be satisfied
throughout the motion, $\dot{\pi^t}(t,r)|_{M_0}=0$ must also be
satisfied for consistency. This condition becomes 
%
\begin{equation}
 \dot{\pi^t}(t,r)|_{M_0} = \{\pi^t(t,r),H(t)\}|_{M_0} = \pi^r{}' + \frac{k_S^2 e^{-(n-2)\phi}}{2f}\mathcal{A}_t
  = 0\ ,
\label{eq:consistency1}
\end{equation}
where
\begin{equation}
 H(t) \equiv \int dr' \mathcal{H}(t,r')\ .
\end{equation}
Thus, we have the following as a secondary constraint,
\begin{equation}
 \mathcal{A}_t = -\frac{2e^{(n-2)\phi}f}{k_S^2}\pi^r{}'\ .
\label{eq:consraint3}
\end{equation}
Above constraint must also be satisfied at all times
and we should require 
%
\begin{equation}
\begin{split}
\left.\frac{\partial}{\partial t}\left(\mathcal{A}_t +
 \frac{2e^{(n-2)\phi}f}{k_S^2}\pi^r{}'\right) \right|_{M_0}
 &= \left.\left\{\mathcal{A}_t +
 \frac{2e^{(n-2)\phi}f}{k_S^2}\pi^r{}',H\right\} \right|_{M_0}\\
 &= \mu -  e^{(n-2)\phi}f(e^{-(n-2)\phi}f\mathcal{A}_r)'= 0\ .
\end{split}
\end{equation}
%
Thus, $\mu$ is determined as
\begin{equation}
 \mu = e^{(n-2)\phi}f(e^{-(n-2)\phi}f\mathcal{A}_r)'\ .
\end{equation}
In a similar way, from $\pi^t{}^\ast= 0$, 
we have also the secondary constraint and expression for $\mu^\ast$ given by 
\begin{align}
 \mathcal{A}_t^\ast &= -\frac{2e^{(n-2)\phi}f}{k_S^2}\pi^r{}^{\prime \ast}\
 ,\label{eq:consraint4}\\
 \mu^\ast &= e^{(n-2)\phi}f(e^{-(n-2)\phi}f\mathcal{A}_r^\ast)'\ .
\end{align}

Since we have the expressions for $\mu$ and $\mu^\ast$, and find out all
constraints, we can find the physical phase space 
\begin{equation}
\begin{split}
M_{\text{phys}}=\bigg\{x\in M \bigg|\pi^t&=\pi^{t\,\ast}=0\ ,\\
 &\mathcal{A}_t =
 -\frac{2e^{(n-2)\phi}f}{k_S^2}\pi^r{}^{\prime}\ , \mathcal{A}_t^\ast
 = -\frac{2e^{(n-2)\phi}f}{k_S^2}\pi^r{}^{\prime \ast}\bigg\}\ .
\end{split}
\end{equation}
Thus, we can compute the Hamiltonian in the physical phase space $M_{\text{phys}}$.
Substituting constraint equations~(\ref{eq:consraint1}), (\ref{eq:consraint2}),
(\ref{eq:consraint3}), (\ref{eq:consraint4}) 
into Hamiltonian~(\ref{eq:H_tot}), we have
\begin{equation}
 \mathcal{H} = 2e^{n\phi}\pi^r \pi^r{}^\ast +
  \frac{2e^{(n-2)\phi}f}{k_S^2}\pi^r{}'\pi^r{}^{\prime \ast} + \frac{k_S^2
  e^{-(n-2)\phi}f}{2}\mathcal{A}_r\mathcal{A}_r^\ast\ ,
\end{equation}
where total derivative terms are omitted. 
Defining new canonical variables
\begin{equation}
 \Phi = \frac{\sqrt{2}}{k_S}\pi^r\ ,\quad
 \Pi = - \frac{k_S}{\sqrt{2}}\mathcal{A}_r\ ,
\end{equation}
%
we can rewrite Hamiltonian~(\ref{eq:H_tot}) as
\begin{equation}
 \mathcal{H} = e^{-(n-2)\phi}f\Pi\,\Pi^\ast + e^{(n-2)\phi}f\Phi'\Phi^{\prime \ast} + k_S^2 e^{n\phi}\Phi\Phi^\ast\ .
\label{eq:H_NH2}
\end{equation}

One can see that Hamiltonian~(\ref{eq:H_NH2}), which is for the scalar modes of the electromagnetic field on $S^n$, is equivalent to that for a complex scalar field in the following.
Let us consider the action of a complex scalar field given by 
\begin{equation}
 S = -\int d^2x \sqrt{-g}\,e^{(n-2)\phi}[\partial_a\chi \partial^a\chi^\ast +
  k_S^2 e^{2\phi}\chi\chi^\ast]\ .
\label{eq:cmscalar}
\end{equation}
From this action, one can easily calculate the Hamiltonian for this complex scalar field as
\begin{equation}
 \mathcal{H} = e^{-(n-2)\phi}f\pi\pi^\ast + e^{(n-2)\phi}f\chi'\chi^{\prime \ast} + k_S^2 e^{n\phi}\chi\chi^\ast\ ,
\end{equation}
where $\pi \equiv e^{(n-2)\phi}\dot{\chi}/f$, a canonical momentum of $\chi^\ast$. 
Since this Hamiltonian is the same expression as Eq.~(\ref{eq:H_NH2}),
it is shown that original action~(\ref{eq:action_scalar2}), which is for the scalar modes of the electromagnetic field, is equivalent to that for a massive complex scalar field, defined by (\ref{eq:cmscalar}).

\subsection{ Near horizon behavior }
\label{subsec:NHB}

Collecting the results in the previous parts of this section, we can write the action for the electromagnetic field on the spherically symmetric spacetimes in the form of
\begin{equation}
\begin{split}
 S=&-\sum_N \int d^2 x \sqrt{-g}\,e^{-(n-2)\phi}[\partial_a \alpha_N
  \partial^a \alpha_N^\ast + (k_V^2 +
  n-1)e^{2\phi}\alpha_N\alpha_N^\ast]\\
  &-\sum_{M\neq 0}\int d^2x \sqrt{-g}\,e^{(n-2)\phi}[\partial_a\chi_M \partial^a\chi_M^\ast +
  k_S^2 e^{2\phi}\chi_M\chi_M^\ast]\ ,
\end{split}
\label{eq:eff_action}
\end{equation}
where summation with respect to $M$ is restored, discarding $l=0$ mode in the scalar sector. 
Action~(\ref{eq:eff_action}) is written in $(t,r)$ coordinates as follows:
\begin{equation}
 \begin{split}
 S=&-\sum_N \int d^2 x \,r^{n-2}[-\frac{1}{f}\partial_t \alpha_N
  \partial_t \alpha_N^\ast + f\partial_r \alpha_N
  \partial_r \alpha_N^\ast + \frac{k_V^2 +
  n-1}{r^2}\alpha_N\alpha_N^\ast]\\
  &-\sum_{M\neq 0}\int d^2x \,r^{-(n-2)}[-\frac{1}{f}\partial_t\chi_M
  \partial_t\chi_M^\ast + f\partial_r\chi_M \partial_r\chi_M^\ast + 
  \frac{k_S^2}{r^2}\chi_M\chi_M^\ast]\ ,
\end{split}
\label{eq:eff_action2}
\end{equation}
where the explicit form of the dilaton, $e^{-2\phi}=r^2$, is used. 
In the near-horizon limit,
$r\rightarrow r_H$, the mass terms in~(\ref{eq:eff_action2}) 
are negligible, and the action takes the form of
\begin{equation}
 \begin{split}
S\simeq
&-r_H^{n-2}\sum_N \int d^2 x \sqrt{-g}\,\partial_a \alpha_N
  \partial^a \alpha_N^\ast
  -r_H^{-(n-2)}\sum_{M\neq 0}\int d^2x \sqrt{-g}\,\partial_a\chi_M \partial^a\chi_M^\ast\ .
\end{split}
\label{eq:2Daction}
\end{equation}

Now, we can see that from action~(\ref{eq:2Daction}), the
electromagnetic field near the horizon in $d$-dimensional spherical
black holes can be described by the theory for an infinite number of 
massless scalar fields on the 2-dimensional spacetime,
whose metric is given by~(\ref{eq:2Dmetric}), as well as the case for a
$d$-dimensional scalar field~\cite{Robinson:2005pd}.
However, the difference between the electromagnetic and scalar fields
appears in the number of degrees of freedom.
That is, there are degeneracies both for scalar and vector sectors in
each $l$-mode, whose numbers of degeneracy are denoted by $D_l(n,0)$ and
$D_l(n,1)$, respectively, and are given by~\cite{Rubin:1984tc}:
\begin{equation}
\begin{split}
D_l(n,0) &= \frac{(2l+n-1)(l+n-2)!}{l!(n-1)!}\ ,\\
D_l(n,1) &= \frac{l(l+n-1)(2l+n-1)(l+n-3)!}{(l+1)!(n-2)!}\ .
\end{split}
\label{eq:Ds}
\end{equation} 
Therefore, each $l$-mode in action~(\ref{eq:2Daction}) contains $D_l(n,0)+D_l(n,1)$ scalar fields. While for the $d$-dimensional scalar field,
no vector mode on $S^n$ exists and
only $D_l(n,0)$ scalar fields are relevant~\cite{Robinson:2005pd}. 
This difference results in the total amount of Hawking
radiation between photons and scalar particles. We will revisit this
point in the next section.

\section{ Hawking radiation of vector particles }
\label{sec:HRandGA}

From the results in the previous section, it suffices to calculate the
Hawking radiation of the scalar fields to know the Hawking radiation of
electromagnetic field.
In this section, we review the derivation of Hawking radiation via the
cancellation of gravitational anomalies for scalar
fields~\cite{Iso:2006ut,Iso:2006xj} in order for this paper to be
self-contained. Then, combining such a  Hawking radiation result for
scalar fields with the results in the previous section, the total amount
of Hawking radiation for the electromagnetic field is estimated.

\subsection{ Hawking radiation of scalar particles as an anomaly eliminator }

In 2-dimensional spacetime, we will regard the horizon as a
boundary of spacetime and discard ingoing modes near the horizon since the ingoing
modes cannot affect the dynamics of the scalar field outside the horizon classically.
As a consequence, such a 2-dimensional theory becomes chiral and a gravitational anomaly might appear.

First, let us split the spacetime into two regions: 
$r_H \leq r \leq r_H + \epsilon$ where the theory is chiral and 
$r_H + \epsilon \leq r$ where the theory is not chiral.
We will take the limit $\epsilon \rightarrow 0$ ultimately.
It is known that the gravitational anomaly arises in 2-dimensional chiral theory and takes the form of \cite{Kimura:1970iv,Alvarez-Gaume:1983ig,Bertlmann:2000da}
%
\begin{equation}
 \nabla_a {T^a}_{b} = -\frac{1}{96\pi \sqrt{-g}}\epsilon^{c
  d}\partial_d \partial_a \Gamma^a_{b c}\ ,
\label{eq:anom}
\end{equation} 
where the convention $\epsilon^{01}=+1$ is used. 
We define $ {N^a}_b $ as
%
\begin{equation}
 \nabla_a {T^a}_{b} \equiv
  \frac{1}{\sqrt{-g}}\partial_a {N^a}_b   \ .
\label{eq:anom2}
\end{equation}
In the non-chiral region $r_H + \epsilon \leq r$, we have 
${N^a}_b = 0$, 
while in the chiral region $r_H \leq r \leq r_H + \epsilon$, 
the components of $ {N^\mu}_\nu $ are
\begin{equation}
 {N^t}_t = {N^r}_r = 0   \ ,\quad
 {N^r}_t = -\frac{1}{192\pi}({f'}^2+f^{\prime \prime }f)   \ ,\quad
 {N^t}_r = \frac{1}{192\pi f^2}(f^{\prime 2}-f^{\prime \prime}f)  \ , 
\end{equation}
where $' \equiv \partial_r$.
In ($t,r$) coordinates, we can write down Eq.~(\ref{eq:anom2}) as
\begin{equation}
\begin{split}
	&\partial_r T_{ (O)~t }^{ ~~~r }
	=0\ ,
	\\
	&\partial_r T_{ (H)~t }^{ ~~~r }
	=
	\partial_r N^{r}_{\;\;t}\ ,
\end{split}
\end{equation}
where the time independence of ${T^a}_b$ is assumed. 
The subscripts, $H$ and $O$, represent the values in the region
$r_H \leq r \leq r_H + \epsilon$ and $r_H + \epsilon \leq r$, 
respectively. These equations can be integrated to give
\begin{equation}
\begin{split}
	T_{ (O)~t }^{ ~~~r }
	=&
	a_O\ ,
	\\
	T_{ (H)~t }^{ ~~~r }
	=&
	a_H
	+ N^{r}_{\;\;t}(r) -  N^{r}_{\;\;t}(r_H)\ ,
\end{split}
\label{eq:T_munu}
\end{equation}
where $ a_O $ and $ a_H $ are integration constants~\cite{Solodukhin:2005ah}.
In particular, note that $a_O$ itself represents the energy flux outside the horizon.

An effective action for the metric $g_{\mu\nu}$,
obtained after integrating out the quantized scalar field, is
\begin{equation}
  W[g_{ab}]=-i \ln \left(\int \mathcal{D} \varphi\,e^{i S[\varphi,\,g_{ab}]}\right)\ ,
\end{equation}
where $S[\varphi,\,g_{ab}]$ is the classical action for one
2-dimensional scalar field.  
By an infinitesimal coordinate transformation in the time direction,
\begin{equation}
 t \rightarrow t - \xi^t(t,r) \ , \quad r \rightarrow r\ ,
\end{equation}
the effective action changes as
\begin{equation}
\begin{split}
	- \delta_\xi W
	=&
	\int d^2 \! x
	\sqrt{ -g } \;
	\xi^t \nabla_a
	\left[
		T_{ (H) ~t }^{ ~~~a } \Sigma_H (r)
		+
		T_{ (O) ~t }^{ ~~~a } \Sigma_O (r)
	\right]
	\\
	=&
	\int d^2 \!x \;
	\xi^{t}
	\left[
		\partial_r
		\left\{
		N^{r}_{\;\;t}
		\Sigma_H\right\}
		+
		\left(
			T_{ (O)~t }^{ ~~~r } - T_{ (H)~t }^{ ~~~r }
			+ N^{ r }_{ \;\;t }
		\right)
		\delta ( r-r_H-\epsilon )
	\right]\ .
	\label{eq:variation1}
\end{split}
\end{equation}
Here, $ \Sigma_O (r) $ and $ \Sigma_H (r) $ are the
 supports of 
$T_{ (O) ~b }^{ ~~~a }$ and 
$T_{ (H) ~b }^{ ~~~a }$, respectively, defined by step function $ \Theta $ as
\begin{eqnarray}
	\Sigma_O (r) \equiv \Theta ( r -r_H -\epsilon )\ ,
	\;\;\;\;\;
	\Sigma_H(r) \equiv 1 - \Theta ( r -r_H -\epsilon )\ .
\end{eqnarray}
Since the first term in the second line of
Eq.~(\ref{eq:variation1}) cannot be canceled by
the delta-function term, it should be canceled by a quantum effect
of the ingoing modes.
The coefficient of the delta function should vanish
to save the diffeomorphism invariance at the quantum level. 
From Eq.~(\ref{eq:T_munu}), this requirement leads to
\begin{equation}
	a_O
	=
	a_H
	-
	N^{r}_{\;\;t} (r_H)\ .
\end{equation}
We need to know $a_H$ to obtain the Hawking flux, $a_O$.
For this purpose, we adopt the boundary condition proposed in \cite{Iso:2006wa}. 
Let us introduce the covariant
energy-momentum tensor $\tilde{T}_{ab}$, which satisfies
a covariant-anomaly equation,
\begin{equation}
 \nabla_a \tilde{T}_{ (H) ~b }^{ ~~~a } = \frac{1}{96\pi
 \sqrt{-g}}\epsilon_{ab} \partial^a R \ .
\label{eq:covanom}
\end{equation}
We impose the vanishing of this
covariant energy-momentum tensor at the horizon since the boundary condition
should be diffeomorphism invariant. In the present case, the covariant
energy-momentum tensor is given by
\begin{eqnarray}
	\tilde{T}_{ (H) ~t }^{ ~~~r }
	=
	T_{ (H) ~t }^{ ~~~r }
	+
	\frac{ 1 }{ 192 \pi }
	\left(
		f f^{ \prime \prime } - 2 f^{ \prime 2 }
	\right).
\end{eqnarray}
The vanishing of this covariant current at the horizon determines $ a_H $ as
\begin{eqnarray}
	a_H
	=\frac{f'(r_H)}{96\pi} = 
	\frac{ \kappa^2 }{ 24\pi }\ .
\end{eqnarray}
Thus, we have
\begin{eqnarray}
	a_O
	=
	\frac{ \kappa^2 }{ 48 \pi }\ .
	\label{eq:aO}
\end{eqnarray}
This is the energy flux in the
outside region, obtained by imposing the cancellation of the gravitational
anomaly at the horizon.
This value exactly coincides with the
energy flux evaluated from a thermal spectrum, 
\begin{equation}
	T^{r}_{\;\;t \; \mathrm{ (thermal) }}
	=
	\int_0^\infty
	\frac{ d\omega }{ 2\pi }
	\frac{\omega}{e^{2\pi \omega/\kappa}-1}
	=
	\frac{ \kappa^2 }{ 48 \pi }\ .
\end{equation}
This result suggests that the Hawking radiation of a scalar field from the spherically
symmetric black holes can be regarded as the anomaly eliminator on horizons.
Combined with the result that the electromagnetic field can be regarded
as a set of an infinite number of massless scalar fields near the horizon,
the above results also suggest that the Hawking radiation of a vector
field should be regarded as the anomaly eliminator on the horizon.

\begin{figure}[h]
\begin{center}
\includegraphics[width=8cm]{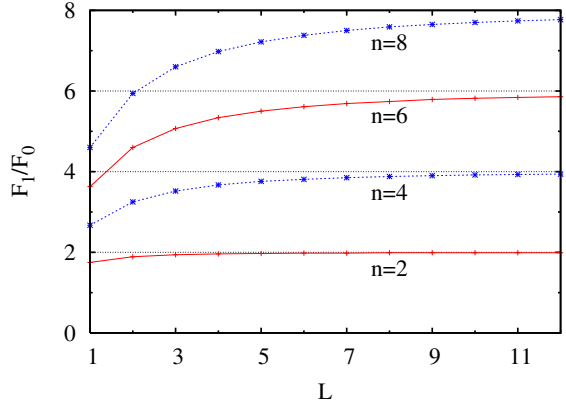}
\caption{
The ratio of Hawking fluxes between a scalar field and an electromagnetic field near the horizon, $F_1(L)/F_0(L)$ (see Eqs.~(\ref{eq:Ds}) and (\ref{eq:fluxes}) for definition), for various spacetime dimensions. We can see that as the cut-off angular momentum $L$ increases, the ratio for each spacetime dimension approaches $n \equiv d-2 $.
\label{fg:ratio}}
\end{center}
\end{figure}

\subsection{ Sum over vector and scalar modes }

Energy flux~(\ref{eq:aO}) is that of the contribution from one scalar field
in 2-dimensional effective action~(\ref{eq:2Daction}).
As mentioned at the last part of Sec.~\ref{subsec:NHB}, however, $D_l(n,0)+D_l(n,1)$ scalar
fields exist in 2-dimensional effective action~(\ref{eq:2Daction}) for
each $l$-mode.

Now, to clarify the quantitative difference in fluxes between scalar
and electromagnetic fields, let us define the following quantities:
\begin{eqnarray}
	&&
	F_0(L) = \sum_{l=0}^{L} D_l(n,0)a_O\ ,
	\nonumber
	\\
	&&
	F_1(L) = \sum_{l=1}^{L} \left[ D_l(n,0)+D_l(n,1) \right]a_O\ .
	\label{eq:fluxes}
\end{eqnarray}
These are the energy fluxes for scalar and electromagnetic fields,
respectively, taking into account the contributions from $ l\leq L $
modes.
The total energy fluxes, obtained by taking limit $ L\to \infty $, diverge for both fields.
Their ratio of $ F_1/F_0 $, however, converges to a finite value depending on the spacetime dimensions:
\begin{equation}
	\lim_{ L \to \infty } \frac{ F_1 (L) }{ F_0 (L) }
	=
	n\ .
	\label{eq:convergence}
\end{equation}
This result is consistent with the fact that $d$-dimensional
electromagnetic fields have $d-2=n$ degrees of freedom. 
In Fig.~\ref{fg:ratio} we show the $L$-dependence of $ F_1(L)/F_0(L) $
for some spacetime dimensions. 
We can say that the
convergence~(\ref{eq:convergence}) is sufficiently rapid especially for
lower dimensions. Finally, note that the above fluxes~(\ref{eq:fluxes})
and their ratio have significance only near the horizon. That is, if
we take into account the scattering by a curvature potential (greybody
factor), the fluxes for higher angular-momentum particles will be
suppressed. Therefore, the total fluxes, obtained in the limit $ L \to
\infty$, will converge for both fields and their ratio will take a
different value from (\ref{eq:convergence}).

\section{Conclusion}
\label{sec:conc}

We have shown that Hawking radiation of vector particles (photons) as
well as a scalar field from the $d$-dimensional spherically symmetric
black holes can be explained from the anomaly cancellation viewpoint. 
This result shows the robustness of the picture, discovered recently and
applied to many black holes in the case of scalar radiation, that
Hawking radiation can be regarded as the anomaly eliminator on horizons.
The spin degree of freedom for the vector field appears in the amount of
total energy flux,
which is larger than that for the scalar field with the dimension-dependent factor of $d-2$.

It will be interesting to generalize the gravitational anomaly method
to graviton/fermion fields.
For the electromagnetic field, it is
essential that the field can be reduced to the 2-dimensional theory for massless scalar fields near the horizon. 
If we can regard the gravitational field and fermion field as a 2-dimensional
massless scalar field and a spinor field near the horizon, respectively,
we will be able to calculate the energy fluxes for these fields from the
anomaly point of view. 
If we succeed in explaining Hawking radiation
for all fundamental fields by the anomaly cancellation method,
our understanding of black hole physics will be significantly enriched.

Although we did not show explicitly that the spectrum of the photon emission
is thermal, the fact that the electromagnetic field can be regarded as
the massless scalar fields near the horizon will enable us to apply the
derivation of the thermal spectrum in the context of 
anomalies~\cite{Iso:2007kt}.
Note also that curvature scattering, which was ignored in this paper, can be
calculated from 2-dimensional action~(\ref{eq:eff_action}).
Therefore, by making use of the thermal spectrum, reproduced in~\cite{Iso:2007kt},
one can calculate the total energy flux observed at infinity from
the anomaly point of view.

Some of recent studies on the counting of black hole entropy 
are also related to anomalies~\cite{Dabholkar:2004yr,Kraus:2005vz,Carlip:2006fm}.
It will be significant to give a unified view both for entropy
and Hawking radiation in the present context.

\section*{Acknowledgements}
We would like to thank Jiro Soda and Keisuke Izumi for useful
discussions. 
K.M. is supported in part by JSPS Grant-in-Aid for Scientific Research,
No.193715 and also by the 21COE program ``Center for Diversity and
Universality in Physics,'' Kyoto University. 
U.M. is supported in part by a grant from the 21st Century COE
Program (Holistic Research and Education Center for Physics
Self-Organization Systems) at Waseda University.

\bibliographystyle{kuma}
\bibliography{EMandGA}

\end{document}